\def\be{\begin{equation}}
\def\ee{\end{equation}}
\def\bea{\begin{eqnarray}}
\def\eea{\end{eqnarray}}
\def\bse{\begin{subequations}}
\def\ese{\end{subequations}}

\documentclass[prl,twocolumn,showpacs,preprintnumbers,amsmath,amssymb]{revtex4}
\usepackage{graphicx}
\usepackage{dcolumn}
\usepackage{bm}
\begin{document}
\preprint{NSF-KITP-05-49}
\title{Nonanalytic Corrections to Fermi-Liquid Behavior in Helimagnets
\vskip 1mm
}
\author{T.R. Kirkpatrick$^{1,2}$ and D. Belitz$^{1,3}$}
\affiliation{$^{1}$Kavli Institute for Theoretical Physics,
University of
California, Santa Barbara, CA 93106\\
             $^{2}$Institute for Physical Science and Technology and Department of
                   Physics, University of Maryland, College Park, MD 20742\\
             $^{3}$Department of Physics and Materials Science Institute, University
                of Oregon, Eugene, OR 97403}
\date{\today}
\begin{abstract}
The Goldstone mode in the ordered phase of itinerant helimagnets, such as MnSi
or FeGe, is determined and shown to have a strongly anisotropic dispersion
relation. The softness of this mode is, in a well-defined sense, in between
that of ferromagnetic and antiferromagnetic magnons, respectively. It is shown
that this soft mode leads to nonanalytic corrections to Fermi-liquid behavior,
with a $T$-contribution to the specific heat coefficient, and a
$T^{5/2}$-contribution to the resistivity. The quasi-particle inelastic
lifetime shows anisotropic behavior in momentum space.
%
\end{abstract}

\pacs{75.30.Ds; 75.30.-m; 75.50.-y; 75.25.+z}

\maketitle

Helimagnets, which display long-ranged helical or spiral spin order, are less
common, and less well known, then their more prevalent ferromagnetic and
antiferromagnetic brethren. They nevertheless have received much attention
lately, due to the unusual properties of the weak itinerant helimagnet MnSi.
This material shows, at ambient pressure, helimagnetic order below a critical
temperature $T_{\text{c}}\approx 30\,{\text K}$ \cite{Ishikawa_et_al_1976}. The
wavelength of the helix is $2\pi/q\approx 180\,{\text{\AA}}$, with $q$ the
pitch wave number. Application of hydrostatic pressure $p$ monotonically
decreases $T_{\text{c}}$ until $T_{\text{c}}$ vanishes at $p = p_{\,\text{c}}
\approx 14\,{\text{kbar}}$ \cite{Pfleiderer_et_al_1997}. In the ordered phase,
neutron scattering shows the helical order with the helix pinned in the
(111)-direction due to crystal-field effects \cite{Ishikawa_Arai_1984}. In the
disordered phase, quasi-static remnants of helical order are still observed at
low temperatures close to the phase boundary, with an approximately isotropic
distribution of the helical axis orientation \cite{Pfleiderer_et_al_2004}. In
the entire disordered phase, up to a temperature of a few Kelvin, and up to the
highest pressures investigated ($\approx 2p_{\,\text{c}}$), pronounced
non-Fermi-liquid behavior of the resistivity is observed, with the temperature
dependence of the resistivity given by $\rho(T\to 0) \propto {\text{const.}} +
T^{3/2}$ \cite{Pfleiderer_Julian_Lonzarich_2001}. In the ordered phase, on the
other hand, the transport behavior is Fermi-liquid-like, with a
$T^2$-dependence of the resistivity. No explanations have been given so far for
these unusual properties. However, it is natural to speculate that the remnants
of helical order that are clearly observed in the paramagnetic phase have
something to do with them.

Given this situation, it is surprising that basic consequences of helical
order, in particular the nature of the Goldstone mode that must result from the
spontaneously broken symmetry, and its effects on observables, are not known.
The present Letter addresses this issue. We will consider an isotropic
itinerant helimagnet in its ordered phase. We will show that there is one
Goldstone mode, the `helimagnon', with a dispersion relation given by
\be
\Omega({\bm k}) \propto \sqrt{k_z^2 + {\bm k}_{\perp}^4/{\bm q}^2}
\label{eq:1}
\ee
in the long-wavelength limit ($\vert{\bm k}\vert < q$). Here $\Omega$ is the
freqency, ${\bm k} = ({\bm k}_{\perp},k_z)$ is the wave vector, and ${\bm q} =
(0,0,q)$ is the pitch or axis vector of the helix, which we take to point in
the $z$-direction. Notice that this dispersion relation is strongly anisotropic
and much softer in the direction transverse to the pitch vector,
$\Omega\propto{\bm k}_{\perp}^2$ for $k_z=0$, than in the $z$-direction,
$\Omega\propto k_z$ for $k_{\perp} = \vert{\bm k}_{\perp}\vert = 0$.

Equation\ (\ref{eq:1}) is our basic result, and all others follow from it. To
the extent that the helimagnon is a well-defined quasi-particle, one expects
its contribution to the internal energy to be $U = \sum_{\bm k} \Omega({\bm
k})\,n(\Omega({\bm k})/T)$. Here $n(x)$ is the Bose distribution function, and
we use unit such that $\hbar = k_{\text B} = 1$. It is easy to see that Eq.\
(\ref{eq:1}) leads to a helimagnon contribution to the specific heat, $C =
\partial U/\partial T$, given by $C\propto T^2$. Adding the leading
Fermi-liquid contribution, which is linear in $T$, the low-temperature specific
heat in the helical phase thus is
\be
C(T\to 0) = \gamma\,T + \gamma_2\,T^2 + O(T^3\ln T).
\label{eq:2}
\ee
Here $\gamma$ is the usual specific-heat coefficient, and $\gamma_2$ depends on
the properties of the helimagnon, see below.

The single-particle Green function is non-diagonal in both spin and momentum
space, but has two well-defined resonances that correspond to a split Fermi
surface. The quasi-particle inelastic scattering rate, or inverse lifetime, due
to scattering by helimagnons, $1/\tau_{\text{in}}({\bm k})$, one expects to be
anisotropic due to the anisotropic nature of the helimagnon spectrum. Indeed,
the poles of the Fermi surface at ${\bm k} = (0,0,k_z)$ turn out to be `hot
spots' with a scattering rate $1/\tau_{\text{in}}\propto T^2$, while elsewhere
the scattering is weaker with $1/\tau_{\text{in}}\propto T^{5/2}$. The
transport lifetime, which determines the resistivity, averages over the Fermi
surface, with the regions with a smaller scattering rate effectively `shunting
out' or `short circuiting' the hot spots \cite{hot_spots_footnote}. As a
result, the temperature dependence of the resistivity at low temperatures is
given by
\be
\rho(T\to 0) = \rho_{\text{imp}} + \rho_2\,T^2 + \rho_{5/2}\,T^{5/2} + O(T^3).
\label{eq:3}
\ee
Here $\rho_{\text{imp}}$ is the residual resistivity due to impurities, the
term proportional to $T^2$ is the usual Fermi-liquid contribution due to the
screened Coulomb interaction between quasi-particles
\cite{Abrikosov_et_al_1963}, and the $T^{5/2}$ contribution is due to
scattering by helimagnons. Notice that the latter, albeit weaker than the
leading Fermi-liquid term, is a nonanalytic function of the temperature.

We now sketch the derivation of these results; a complete account of the theory
will be given elsewhere \cite{us_tbp}. Our starting point is a microscopic
action $S_0$ for free or band electrons, and an electron-electron interaction
$S_{\text{int}}$. The spin-singlet interaction will not be crucial for our
purposes, and we suppress it for simplicity. The spin-triplet interaction has
an isotropic, short-ranged contribution that can be modelled by the usual
point-like Stoner interaction with an interaction amplitude $\Gamma_{\text t}$
\cite{Hertz_1976}. In materials whose crystal structure lacks inversion
symmetry, the spin-orbit interaction leads to an additional chiral term
\cite{Dzyaloshinski_1958, Moriya_1960}. This Dzyaloshinski-Moriya (DM)
interaction is believed to be responsible for the helimagnetism observed in
MnSi and FeGe \cite{Bak_Jensen_1980}. The spin-triplet interaction term can
then be written
\bse
\label{eqs:4}
\be
S_{\text{int}}^{\text t} = \int d{\bm x}\,d{\bm y}\int_0^{1/T} d\tau\
   n_{\text s}^i({\bm x},\tau)\,A_{ij}({\bm x}-{\bm y})\,n_{\text s}^j({\bm
   y},\tau).
   \label{eq:4a}
\ee
Here $n_{\text s}^i$ ($i=1,2,3$) are the components of the electronic spin
density field which depend on position ${\bm x}$ and imaginary time $\tau$, and
\be
A_{ij}({\bm x}-{\bm y}) = \delta_{ij}\,\Gamma_{\text t}\,\delta({\bm x}-{\bm
y}) + \epsilon_{ijk}\,C_k({\bm x}-{\bm y}),
\label{eq:4b}
\ee
\ese
with $\epsilon_{ijk}$ the Levi-Civita tensor and the $C_k$ the components of a
vector-valued function of the position. In Fourier space, $C_k({\bm p}\to 0) =
c\,p_{\,k}$. The coefficient $c$ plays the role of a chiral coupling constant
in the theory.

A Hubbard-Stratonovich transformation that decouples the interaction term along
the lines of Ref.\ \cite{Hertz_1976} leads to a Landau-Ginzburg-Wilson (LGW)
functional for a helimagnet:
\bse
\label{eqs:5}
\be
\hskip -0pt S_{\text{heli}}[{\bm M}] = S_{\text{fm}}[{\bm M}] + c\int dx\, {\bm
M}(x)\cdot\left[{\bm\nabla}\times{\bm M}(x)\right],
\label{eq:5a}
\ee
\vskip -5mm
\bea
S_{\text{fm}}[{\bm M}] &=& \frac{1}{2} \int dx\,dy\ {\bm
M}(x)\,\Gamma(x-y)\,{\bm M}(y)
\nonumber\\
 && +\, \frac{u}{4}\int dx\, \left({\bm M}^2(x)\right)^2.
\label{eq:5b}
\eea
The two-point vertex $\Gamma$ reads, in Fourier space,
\be
\Gamma({\bm p},i\Omega_n) = t + a\,{\bm p}^2 + b\,\vert\Omega_n\vert/\vert{\bm
p}\,\vert.
\label{eq:5c}
\ee
\ese
Here we use a four-vector notation, $x\equiv ({\bm x},\tau)$ and $\int dx
\equiv \int_V d{\bm x}\,\int_0^{1/T}$. $\Omega_n = 2\pi Tn$ denotes a bosonic
Matsubara frequency, and ${\bm M}$ is the order parameter or
Hubbard-Stratonovich field with components $M_i$ ($i=1,2,3$), whose expectation
value is proportional to the magnetization. $t$, $a$, $b$, and $u$ are the
parameters of the LGW theory, they can be expressed in terms of the parameters
of the underlying microscopic theory if desirable. $S_{\text{fm}}$ is Hertz's
action for a quantum itinerant ferromagnet \cite{Hertz_1976,
nonanalyticity_footnote}.

The action given by Eqs.\ (\ref{eqs:5}) is easily seen to posses a helical
saddle-point solution
\be
M_{\text{sp}}(x) = m\,\left(\cos(qz),\sin(qz),0\right).
\label{eq:6}
\ee
The free energy in saddle-point approximation is minimized by $q=c/2a$, and the
amplitude $m$ is determined by the saddle-point equation of state
\be
1 = c\,q - \frac{1}{V}\sum_{\bm p} T\sum_{i\omega_n}
    \frac{2\Gamma_{\text t}}{G_0^{-1}({\bm p},i\omega_n)\,
            G_0^{-1}({\bm p}-{\bm q},i\omega_n) - \lambda^2}.
\label{eq:7}
\ee
Here $\lambda = m\Gamma_{\text t}$, and
\be
G_0^{-1}({\bm p},i\omega_n) = i\omega_n - \xi_{\bm p}
\label{eq:8}
\ee
is the inverse Green function for noninteracting electrons described by the
action $S_0$. $\xi_{\bm p} = \epsilon_{\bm p} - \mu$ with $\epsilon_{\bm p}$
the energy-momentum relation and $\mu$ the chemical potential, and $\omega_n =
2\pi T (n+1/2)$ is a fermionic Matsubara frequency. For explicit calculations
we will use an effective-mass approximation, $\epsilon_{\bm p} = {\bm
p}^2/2m_{\text e}$, with $m_{\text e}$ the electronic effective mass.

\begin{figure}[t]
\vskip -0mm
\includegraphics[width=7.0cm]{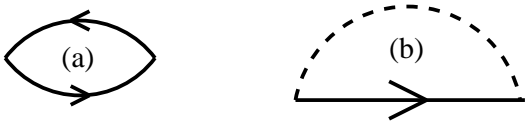}
\caption{(a) The reference ensemble spin susceptibility, and (b) the
lowest-order self-energy diagram contributing to the inelastic scattering rate.
Solid and dashed line denotes the reference ensemble Green function and the
helimagnon, respectively.}
\label{fig:1}
\end{figure}

We now expand the action about the saddle-point solution, writing ${\bm M}(x) =
M_{\text{sp}}({\bm x}) + \delta M(x)$. To Gaussian order in the fluctuations
$\delta{\bm M}$ one obtains
\bea
S_{\text{heli}} &=& S_{\text{sp}} + \frac{1}{2}\int dx\,dy\ \delta
M_i(x)\,\Bigl[\delta_{ij}\,\delta(x-y)\,\Gamma_{\text t}
\nonumber\\
&& \hskip -40pt - \epsilon_{ijk}\,\delta(x-y)\,\Gamma_{\text
t}\,c\,\frac{\partial}{\partial x_k} -
\chi_{\text{ref}}^{ij}(x,y)\,\Gamma_{\text t}^2\Bigr]\,\delta M_j(y).
\label{eq:9}
\eea
Here $S_{\text{sp}}$ is the saddle-point action, and
\be
\chi_{\text{ref}}^{ij}(x,y) = \langle n_{\text s}^i(x)\,n_{\text
s}^j(y)\rangle_{\text{ref}}^{\text c}
\label{eq:10}
\ee
is a spin susceptibility for electrons in a reference ensemble described by an
action for noninteracting electrons in an external field given by
$\Gamma_{\text t}{\bm M}_{\text{sp}}({\bm x})$.
\be
S_{\text{ref}} = S_0 -\Gamma_{\text t}\int dx\ {\bm M}_{\text{sp}}({\bm
x})\cdot {\bm n}_{\text{s}}(x).
\label{eq:11}
\ee
The Green function associated with this reference ensemble takes the form
\bse
\label{eqs:11'}
\bea
G_{{\bm k}{\bm p}}(i\omega_n) &=& \delta_{{\bm k}{\bm
p}}\,\left[\sigma_{+-}\,a_+({\bm k},{\bm q};i\omega_n) +
            \sigma_{-+}\,a_-({\bm k},{\bm q};i\omega_n)\right]
\nonumber\\
&& \hskip -60pt + \delta_{{\bm k}+{\bm q},{\bm p}}\,\sigma_+\,b_+({\bm k},{\bm
q};i\omega_n)
   + \delta_{{\bm k}-{\bm q},{\bm p}}\,\sigma_-\,b_-({\bm k},{\bm
   q};i\omega_n).
\label{eq:11'a}
\eea
Here $\sigma_{\pm} = (\sigma_1 \pm i\sigma_2)/2$ and $\sigma_{+-} =
\sigma_+\sigma_-$, $\sigma_{-+} = \sigma_-\sigma_+$, with $\sigma_{1,2}$ Pauli
matrices, and
\bea
a_{\pm}({\bm k},{\bm q};i\omega_n) &=& G_0^{-1}({\bm k}\pm{\bm
q},i\omega_n)/N_{\pm}({\bm k},{\bm q};i\omega_n),\hskip 20pt
\label{eq:11'b}\\
b_{\pm}({\bm k},{\bm q};i\omega_n) &=& \lambda/N_{\pm}({\bm k},{\bm
q};i\omega_n),
\label{eq:11'c}
\eea
with
\be
N_{\pm}({\bm k},{\bm q};i\omega_n) = G_0^{-1}({\bm k},i\omega_n)\,G_0^{-1}({\bm
k}\pm{\bm q},i\omega_n) - \lambda^2.
\label{eq:11'd}
\ee
\ese
$\langle \ldots\rangle_{\text{ref}}^{\text c}$ in Eq.\ (\ref{eq:10}) denotes a
connected correlation function with respect to $S_{\text{ref}}$.
$\chi_{\text{ref}}$ is given in terms of the reference ensemble Green function
as shown in Fig.\ \ref{fig:1}.

The coefficients of the quadratic form given by Eq.\ (\ref{eq:9}) are thus
known explicitly, and one can diagonalize it to find the excitations of the
helically ordered state. We expect one soft or massless excitation, which
corresponds to the Goldstone mode associated with the spontaneously broken
symmetry in the ordered state. To find this mode, we parameterize the order
parameter as follows \cite{soft_mode_footnote}
\be
{\bm M}(x) = m\,\left(\begin{array}{c} \cos({\bm q}\cdot{\bm x} + \phi(x)) \\
                                       \sin({\bm q}\cdot{\bm x} + \phi(x)) \\
                                       \psi_1(x)\sin({\bm q}\cdot{\bm x})
                                          + \psi_2(x)\cos({\bm q}\cdot{\bm x})
                      \end{array}\right),
\label{eq:12}
\ee
and expand to first order in $\phi$. The diagonalization problem is now
straightforward, although very cumbersome. The Goldstone mode, whose softness
is ensured by the equation of state (\ref{eq:7}) is given by the linear
combination
\bse
\label{eqs:13}
\bea
g(k) = \phi({\bm k},i\Omega_n) &-& i(k_y/q)\,[1 + O({\bm
k}_{\perp}^2)]\,\psi_1(k) \nonumber\\
 &&\hskip -20pt - i(k_x/q)\,[1 + O({\bm k}_{\perp}^2)]\,\psi_2(k),
\label{eq:13a}
\eea
where $k\equiv({\bm k},i\Omega_n)$. The helimagnon is the $g$-$g$ correlation
function, which for small frequencies ($\vert\Omega_n\vert\ll\lambda$) and
small wave numbers ($\vert{\bm k}\vert\ll q$) reads
\bea
\langle g(k)\,g(-k)\rangle &=& \frac{1}{2N_{\text
F}\lambda^2}\,\frac{1}{2f({\bm k})}\,\biggl(\frac{1}{f({\bm k}) +
i\Omega_n/2\lambda}
\nonumber\\
&& \hskip 30pt + \frac{1}{f({\bm k}) - i\Omega_n/2\lambda}\biggr),
\label{eq:13b}
\eea
where
\be
f({\bm k}) = \sqrt{\frac{1}{3}\,k_z^2/(2k_{\text F})^2 + \frac{1}{6}\,{\bm
k}_{\perp}^4/(2qk_{\text F})^2}
\label{eq:13c}
\ee
\ese
with $k_{\text F} = \sqrt{2m_{\text e}\mu}$ the Fermi wave number and $N_{\text
F} = k_{\text F}m_{\text e}/2\pi^2$. We see that the helimagnon is a
propagating excitation with a dispersion relation given by Eq.\ (\ref{eq:1}).

We now turn to the effect of this soft mode on various observables. The
specific heat can be calculated either by the elementary method mentioned in
the context of Eq.\ (\ref{eq:2}), or by calculating the free energy in
saddle-point approximation. Either way one obtains
\be
C_V = \frac{9\,\zeta(3)}{2\sqrt{2}\pi}\,k_{\text F}^3\,\frac{q}{k_{\text F}}\,
\frac{T^2}{\lambda^2},
\label{eq:14}
\ee
where $\zeta$ denotes the Riemann zeta function.

The quasi-particle relaxation rate can be extracted from the imaginary part of
the electronic self energy in the usual way. To linear order, the helimagnon
contribution to the latter is given by the diagram shown in Fig.\ \ref{fig:1}.
For a quasiparticle on the Fermi surface, and for ${\bm k} = ({\bm
k}_{\perp}=0,k_z)$, one finds a scattering rate at asymptotically low
temperatures
\bse
\label{eqs:15}
\bea
\frac{1}{\tau_{\text F}({\bm k})} &=& \frac{9\sqrt{2}k_{\text F}^4 q^3}{64
N_{\text F} m_{\text e}^2}\,\frac{1}{(\xi_{\bm k} + \xi_{{\bm k}+{\bm q}})^4}\,
\left(\frac{(2-Q)^2}{[K_z + (2-Q)^2]^{3/2}}\right.
\nonumber\\
&& + \left.\frac{(2+Q)^2}{[K_z + (2+Q)^2]^{3/2}}\right)\,T^2.
\label{eq:15a}
\eea
Here $K_z = \sqrt{3}k_{\text F}k_z/m_e\lambda$ and $Q = \sqrt{6}k_{\text
F}q/2m_e\lambda$. For ${\bm k}_{\perp}\neq 0$ the asymptotic scattering rate is
smaller,
\be
\frac{1}{\tau_{\text F}({\bm k})} = \frac{9\sqrt{2}A}{4\pi^3}\,\frac{k_{\text
F}^4 q^3}{N_{\text F}m_{\text e}^2\vert{\bm k}_{\perp}\vert\sqrt{\lambda}}\,
\frac{1}{(\xi_{\bm k} + \xi_{{\bm k}+{\bm q}})^4}\,T^{5/2},
\label{eq:15b}
\ee
with $A = 5.13\ldots$.
\ese
The crossover between these two types of behavior occurs at a temperature
proportional to $T_{\times} = (q/k_{\text F})(k_{\perp}^2/k_z^2)\,\lambda$.

To determine the temperature dependence of the resistivity we adapt a simple
argument given for antiferromagnets by Rosch in Ref.\ \cite{Rosch_1999}, who
also showed that a variational solution of the Boltzmann equation gives
qualitatively the same answer. Let $T_0$ and $\tau_0$ be microscopic
temperature and time scales, respectively, $t = T/T_0$, and $\tau_{\text{imp}}$
the mean-free time due to impurity scattering. From Matthiessen's rule one has
on the `cold' parts of the Fermi surface, where the scattering rate goes like
$T^{5/2}$, $1/\tau_{\text{cold}} = 1/\tau_{\text{imp}} + t^{5/2}/\tau_0$. At
the `hot spots' one has $1/\tau_{\text{hot}} = 1/\tau_{\text{imp}} +
\alpha\,t^2/\tau_0$, with $\alpha$ a coefficient. From the dependence of the
crossover temperature $T_{\times}$ on $k_z$ and $k_{\perp}$, and applying
elementary geometry (for small $q$ the Fermi surface is approximately
spherical) one finds that the effective area of the `hot spots', normalized by
the area of the Fermi surface, is linear in $T$, $a_{\text{hot}} = \beta\,t$,
with $\beta$ another coefficient. The hot and cold areas of the Fermi surface
are expected to act effectively like parallel resistors, i.e., the resistivity
will be given by
\bse
\label{eqs:16}
\be
\rho = \frac{m_{\text e}}{n\,e^2}\,\left[a_{\text{hot}}\,\tau_{\text{hot}} +
(1-a_{\text{hot}})\,\tau_{\text{cold}}\right]^{-1},
\label{eq:16}
\ee
with $n$ and $e$ the electron density and charge, respectively. For the
low-temperature behavior one thus finds
\be
\rho = \rho_{\text{imp}}\,\left(1 + t^{5/2}\tau_{\text{imp}}/\tau_0 +
O(t^3)\right)
\label{eq:16b}
\ee
\ese
with $\rho_{\text{imp}} = m_{\text e}/n\,e^2\tau_{\text{imp}}$ the residual
resistivity and $\rho_{0} = m_{\text e}/n\,e^2\tau_{0 }$. This expression
remains valid in the clean limit, when $\tau_{\text{imp}}\to\infty$. Adding the
usual Fermi-liquid contribution one obtains Eq.\ (\ref{eq:3}).

We finally discuss these results. The anisotropic nature of the dispersion
relation, Eq.\ (\ref{eq:1}), is analogous to that of the Goldstone mode in
cholesteric liquid crystals \cite{Lubensky_1972}, which also display helical
order, although with a slightly different order parameter than a helimagnet. In
either case, the reason for the anisotropy is rotational invariance, which can
be seen as follows \cite{Chaikin_Lubensky_1995}: A simple phase fluctuation,
with $\psi_1 = \psi_2 = 0$ in Eq.\ (\ref{eq:12}), would lead to a soft mode
(the phase $\phi$) with $\Omega^2({\bm k}) \propto {\bm k}^2$. However, an
overall tilt of the helical axis, which cannot cost any energy, can be
accomplished by choosing a phase $\phi(x) = \phi({\bm x}) = \varphi_1\,x +
\varphi_2\,y$. To avoid this problem, the soft-mode action must not depend on
${\bm\nabla}_{\perp}\phi$, and the lowest order of ${\bm k}_{\perp}$ allowed in
the dispersion relation is ${\bm k}_{\perp}^4$.

The helimagnon dispersion relation, Eqs.\ (\ref{eq:1}) and (\ref{eqs:13}), is
intermediate between isotropic ferromagnets ($\Omega\propto{\bm k}^2$) and
antiferromagnets ($\Omega\propto\vert{\bm k}\vert$), respectively. In agreement
with this observation, the contribution to the specific heat is in between the
one from ferromagnetic ($C\propto T^{3/2}$) and antiferromagnetic ($C\propto
T^3$) magnons \cite{Kittel_1963}. Although it is weaker than the leading
Fermi-liquid contribution, it is distinct from any contribution within
Fermi-liquid theory, which does not produce a $T^2$-term
\cite{Abrikosov_et_al_1963}. Analogously, the helimagnon contribution to the
temperature dependence of the resistivity is intermediate between the one from
ferromagnetic magnons ($T^2$, Ref.\ \onlinecite{Ueda_Moriya_1975}) and
antiferromagnetic magnons ($T^5$, Ref.\ \onlinecite{Ueda_1977}). Here the
effect of the helimagnons is even more striking, as they produce a nonanalytic
temperature dependence of the resistivity. Since the helimagnon contribution is
stronger than the one from phonons, this effect should be observable if the
$T^2$ background can be subtracted \cite{mass_footnote}.

All of these results are consistent with the experimental observation that the
leading low-temperature behavior in the ordered phase of MnSi is consistent
with Fermi-liquid theory. We point out, however, that the result for the
specific heat is more general and robust than the one for the resistivity. The
former depends only on the existence of well-defined quasi-particles with a
dispersion relation given by Eq.\ (\ref{eq:1}), whereas the latter corresponds
to just a lowest-order variational solution of the Boltzmann equation. A more
sophisticated transport theory, which takes into account either mode-mode
coupling effects, or an interplay between quenched disorder and helimagnons
beyond Matthiessen's rule, might well change the leading temperature dependence
of the resistivity. While the Boltzmann result seems to suggest that the
non-Fermi-liquid behavior observed in the disordered phase is unlikely to be
due to partial or quasi-static helical order, more work is needed to test this
conclusion.

We thank the participants of the Workshop on Quantum Phase Transitions at the
KITP at UCSB, and in particular Christian Pfleiderer and Thomas Vojta, for
stimulating discussions. This work was supported by the NSF under grant Nos.
DMR-01-32555, DMR-01-32726, and PHY-99-07949.


\begin{thebibliography}{25}
\bibitem[{sof()}]{soft_mode_footnote}
\bibinfo{note}{While this is not the most general parameterization of ${\bm
  M}(x)$, it suffices to capture the Goldstone mode. All additional degrees of
  freedom can be shown to be massive.}
\bibitem[{mas()}]{mass_footnote}
\bibinfo{note}{One needs to keep in mind that our results are for isotropic
  helimagnets with no preferred direction for the helical axis. In real
  materials, crystal-field effects lead to a mass of the Goldstone mode, which
  alters the behavior at asymptotically low temperature. The same is true, for
  instance, for Heisenberg ferromagnets. Since ferromagnetic magnons, and their
  effects on the thermodynamic and transport properties, are readily observable
  in the latter, one expects the same to be true for helimagnets, at least in
  materials with a weak pinning of the helical axis.}
\end{thebibliography}

\end{document}